\documentclass[twocolumn,prl,floatfix,superscriptaddress,nobibnotes]{revtex4-1}

\usepackage{amsmath}
\usepackage{amssymb}
\usepackage{graphicx}
\usepackage{color}

\begin{document}

\title{Flux-induced Majorana modes in full-shell nanowires}

\author{S.~Vaitiek\.{e}nas}
\affiliation{Center for Quantum Devices and Microsoft Quantum Lab--Copenhagen, Niels Bohr Institute, University of Copenhagen, 2100 Copenhagen, Denmark}
\author{M.-T.~Deng}
\affiliation{Center for Quantum Devices and Microsoft Quantum Lab--Copenhagen, Niels Bohr Institute, University of Copenhagen, 2100 Copenhagen, Denmark}
\author{P.~Krogstrup}
\affiliation{Center for Quantum Devices and Microsoft Quantum Lab--Copenhagen, Niels Bohr Institute, University of Copenhagen, 2100 Copenhagen, Denmark}
\author{C.~M.~Marcus}
\affiliation{Center for Quantum Devices and Microsoft Quantum Lab--Copenhagen, Niels Bohr Institute, University of Copenhagen, 2100 Copenhagen, Denmark}

\date{\today}

\begin{abstract}
We demonstrate a novel means of creating Majorana zero modes using magnetic flux applied to a full superconducting shell surrounding a semiconducting nanowire core, unifying approaches based on proximitized nanowires and vortices in topological superconductors. In the destructive Little-Parks regime, reentrant regions of superconductivity are associated with integer number of phase windings in the shell. Tunneling into the core reveals a hard induced gap near zero applied flux, corresponding to zero phase winding, and a gapped region with a discrete zero-energy state for flux around $\Phi_{0} = h/2e$, corresponding to $2 \pi$ phase winding.  Coulomb peak spacing in full-shell islands around one applied flux shows exponentially decreasing deviation from $1e$ periodicity with device length, consistent with the picture of Majorana modes located at the ends of the wire.
\end{abstract}

\maketitle

Majorana zero modes (MZMs) at the ends of one-dimensional topological superconductors are expected to exhibit braiding statistics \cite{Kitaev2003,Read2000}, opening a path toward topologically protected quantum computing \cite{Nayak2008,DasSarma2015}. Among the proposals to realize MZMs, one approach \cite{Lutchyn2010,Oreg2010}  based on semiconductor nanowires with strong spin-orbit coupling subject to a Zeeman field and superconducting proximity effect has received particular attention, yielding numerous compelling experimental signatures \cite{Mourik2012,Deng2016,Zhang2018,Albrecht2016,Lutchyn2018}. An alternative route to MZMs aims to create vortices in spinless superconductors, by various means, for instance by coupling a vortex in a conventional superconductor to a topological insulator \cite{Fu2008,Xu2015}, using doped topological insulators \cite{Hosur2011,Wang2018}, or using vortices in exotic quantum Hall analogs of spinless superconductors \cite{DasSarma2005}.

\begin{figure}[h!b]
\includegraphics[width=\linewidth]{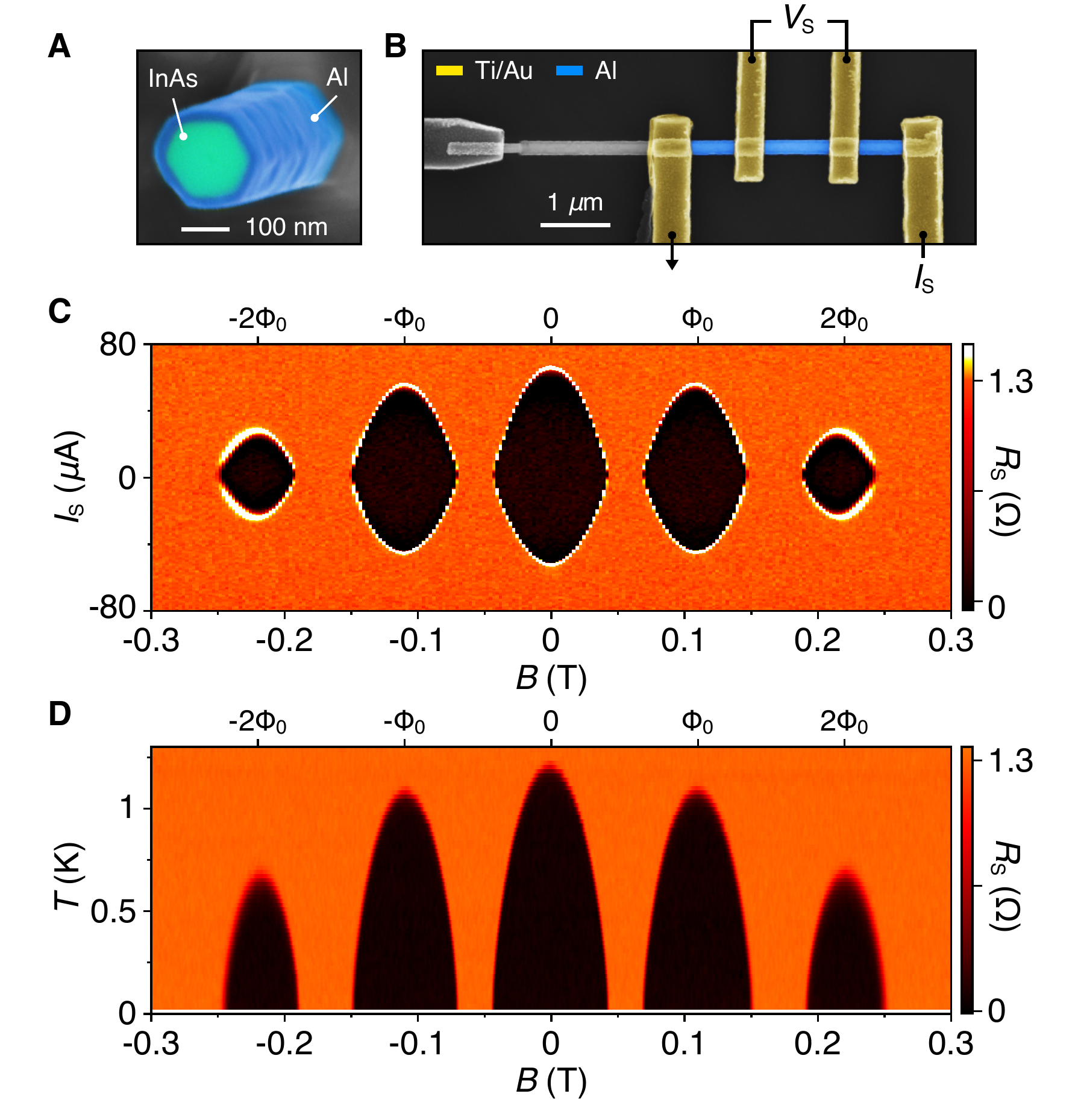}
\caption{\label{fig:1} \textbf{Destructive Little-Parks regime in full-shell nanowire.} (\textbf{A}) Colorized material-sensitive  electron micrograph of InAs-Al hybrid nanowire. Hexagonal InAs core (maximum diameter 130~nm) with 30~nm full-shell epitaxial Al. (\textbf{B}) Micrograph of device 1, colorized to highlight 4-probe measurement setup. (\textbf{C}) Differential resistance of the Al shell, $R_{\rm s}$, as a function of current bias, $I_{\rm s}$, and axial magnetic field, $B$, measured at $20$~mK. Top axis shows flux, $BA_{\rm wire}$, in units of the flux quantum $\Phi_{0} = h/2e$. Superconducting lobes are separated by destructive regions near odd half-integer flux quanta. (\textbf{D}) Temperature evolution of $R_{\rm s}$ as a function of $B$ measured around $I_{\rm s}=0$. Note that $R_{\rm s}$ equals the normal-state resistance in all destructive regimes.}
\end{figure}

The approach demonstrated in this paper, based on superconducting phase winding in an Al shell surrounding an InAs nanowire core, contains elements of both the Lutchyn-Oreg scheme \cite{Lutchyn2010,Oreg2010} and vortex-based schemes \cite{Fu2008} for creating MZMs. In the destructive Little-Parks regime \cite{Little1962,deGennes1981}, the modulation of critical current and temperature with flux applied along the hybrid nanowire results in reentrant superconductivity \cite{Liu2001,Sternfeld2011} where each region is associated with a quantized number of twists of the superconducting phase \cite{Tinkham1966}. The result is a series of topologically locked boundary conditions for the proximity effect of the core, where the number of phase twists in the Al shell corresponds to the number of phase vortices in the nanowire core \cite{Footnote1}. 

We observe that tunneling into the core in the zeroth superconducting lobe, around zero flux, yields a hard proximity-induced gap with no subgap features. In the superconducting regions around one quantum of applied flux, corresponding to phase twists of $\pm 2\pi$ in the shell, tunneling spectra into the core shows stable zero-bias peaks, indicating a discrete subgap state fixed at zero energy, consistent with the Majorana picture. Further support for this interpretation, based on full-shell Coulomb islands, is then presented. We find that in the zeroth lobe, Coulomb blockade conductance peaks show $2e$ spacing, indicating Cooper-pair tunneling and an induced gap exceeding the island charging energy. In the first lobe, peak spacings are roughly $1e$-periodic, with slight even-odd alternation that vanishes exponentially with island length, suggesting overlapping Majorana modes at the two ends of the Coulomb island, as investigated previously \cite{Albrecht2016,vanHeck2016}. These experimental observations are consistent with the recent theory \cite{Lutchyn2018_2} showing that a radial Rashba field arising from the band bending at the semiconductor-superconductor interface \cite{Antipov2018,Mikkelsen2018}, along with an odd multiple of $2\pi$ phase twists in the boundary condition, can induce a topological state with MZMs.

Phase winding in the full-shell geometry represents the continuum limit of discrete boundaries with differing phases. Phase control of Andreev bound states was investigated experimentally for two superconductors as a function of phase difference in Refs.~\cite {Pillet2010,Chang2013,Bretheau2017}. In the context of topological states, Altland and Zirnbauer considered two  superconducting boundaries with phase difference of $\pi$ in their original study of symmetry classes of Andreev billiards \cite{Altland1997}. Phase difference between superconducting boundaries was shown theoretically to influence the topological transition and the appearance of MZMs in planar Josephson junctions \cite{Hell2017,Pientka2017} as well as nanowire models \cite{Kotetes2015,Hansen2016,Stanescu2018}. Control of topological states by multiple phase differences was investigated in Refs.~\cite{vanHeck2014,Riwar2016}. A unique feature of the continuous superconducting shell is the rigidity of phase winding by fluxoid quantization \cite{Tinkham1966}. In this case, a topologically constrained boundary condition locks the topological phase within.

\begin{figure}[t]
\includegraphics[width=\linewidth]{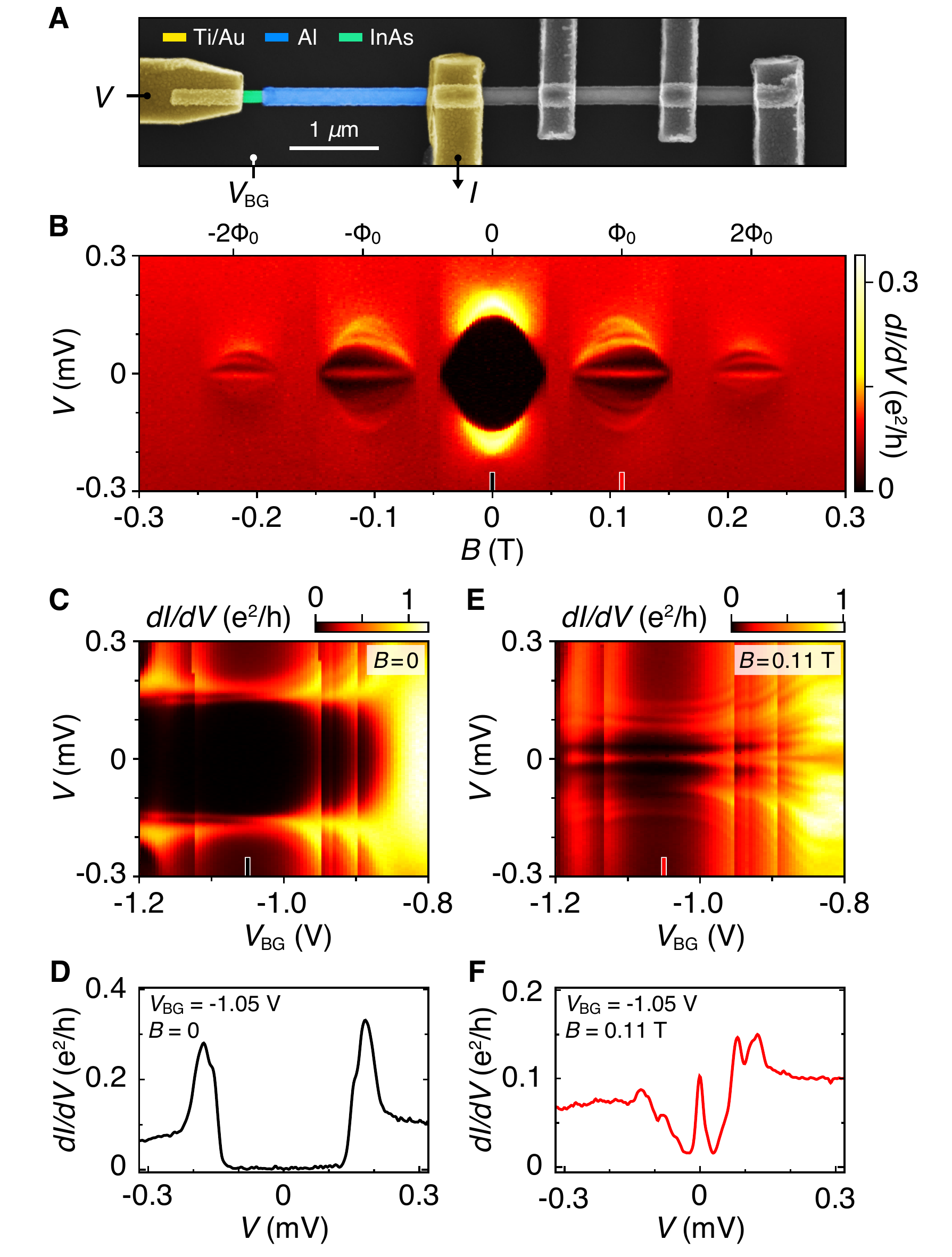}
\caption{\label{fig:2} \textbf{Tunneling spectrum: hard gap in the zeroth lobe, zero-bias peak in the first lobe.} (\textbf{A}) Micrograph of device 1 colorized to highlight tunneling spectroscopy set-up. (\textbf{B}) Differential conductance, $dI/dV$, as a function of source-drain bias voltage, $V$, and axial field, $B$. The zeroth lobe shows a hard superconducting gap, the first lobes show zero-bias peak, the second lobes show non-zero subgap states. The lobes are separated by featureless normal-state spectra. (\textbf{C}) Zero-field conductance as a function of $V$ and back-gate voltage, $V_{\rm BG}$. (\textbf{D}) Line-cut of the conductance taken at $B=0$ and $V_{\rm BG}=-1.05$~V. (\textbf{E} and \textbf{F}) Similar to (C) and (D), measured in the first lobe at $B=110$~mT.
}
\end{figure}

InAs nanowires with wurtzite crystal structure were grown along the [0001] direction by the vapor-liquid-solid method using molecular beam epitaxy (MBE). The nanowires have a hexagonal cross section with maximum diameter ${\rm D}=130$~nm. A 30~nm epitaxial Al layer was grown while rotating the sample, yielding a fully enclosing shell (Fig.~1A) \cite{Krogstrup2015}. Wires were placed on a doped Si substrate capped with thermal oxide. The Al shell was lithographically patterned and selectively etched. Ti/Au ohmic contacts were patterned and deposited following Ar-ion milling. For some devices, Ti/Au side gates were patterned in a subsequent lithographic step (Fig.~3A). Standard ac lock-in measurements were carried out in a dilution refrigerator with a base temperature of 20~mK. Magnetic field was applied parallel to the nanowire using three-axis vector magnet.  Two device geometries, measured in three devices each, showed similar results. Data from two devices are presented: device 1 was used for 4-probe measurements of the shell (Fig.~1B) and tunneling spectroscopy of the core (Fig.~2A); device 2 comprised six Coulomb islands of different lengths fabricated on a single nanowire, each with separate ohmic contacts, two side gates to trim tunnel barriers, and a plunger gate to change occupancy (Fig.~3A).

Differential resistance of the shell, $R_{\rm s} = dV_{\rm s}/dI_{\rm s}$, measured for device 1 as a function of bias current, $I_{\rm s}$, and axial magnetic field, $B$, showed a lobe pattern characteristic of the destructive regime (Fig.~1C) with maximum switching current of $70~\mu$A at $B = 0$, the center of the zeroth lobe. Between the zeroth and first lobes, supercurrent vanished at $\vert B\vert=45$~mT, re-emerged at $70$~mT, and had a maximum near the center of the first lobe, at $\vert B\vert =110$ mT. A second lobe with smaller critical current was also observed, but no third lobe.

Temperature dependence of $R_{\rm s}$ around zero bias yielded a reentrant phase diagram with superconducting regions separated by destructive regions with temperature-independent normal-state resistance $R_{\rm s}^{\rm (N)} = 1.3~\Omega$ (Fig.~1D). $R_{\rm s}^{\rm (N)}$ and shell dimensions from Fig.~1A yield a Drude mean free path of $l=19$~nm. The dirty-limit shell coherence length $\xi_{\rm S}=\sqrt[]{\pi\hbar v_{\rm F} l/24 k_{\rm B} T_{\rm C}}$ \cite{Tinkham1966,Gordon1984} can then be found using the zero-field critical temperature $T_{\rm C}=1.2$~K from Fig.~1D and Fermi velocity of Al, $v_{\rm F}=2\times 10^6$~m/s \cite{Kittel2005}, with Planck constant $\hbar$ and Boltzmann constant $k_{\rm B}$, yielding $\xi_{\rm s} = 180$~nm. The same value for $\xi_{\rm S}$ is found using the onset of the first destructive regime \cite{Schwiete2009}.

Differential conductance, $dI/dV$, as a function of source-drain voltage, $V$, measured in the tunneling regime as a probe of the local density of states at the end of the nanowire is shown in Fig.~2. The Al shell was removed at the end of the wire and the tunnel barrier was controlled by the global back-gate at voltage $V_{\rm BG}$. At zero field, a hard superconducting gap was observed throughout the zeroth superconducting lobe (Fig.~2,~B~and~D). Similar to the supercurrent measurements presented above, the superconducting gap in the core closed at $|B|=55$~mT and reopened at $65$~mT, separated by a gapless destructive regime. Upon reopening, a narrow zero-bias conductance peak was observed throughout the first gapped lobe (Fig.~2,~B~and~F). Several subgap states separated from the zero-bias peak were also visible in the first lobe, as discussed in detail below. The asymmetry of the subgap features around one flux quantum reflects a competition between diamagnetic flux response of the shell and the core. The first lobe persist to $\pm 150$~mT, above which a second gapless destructive regime was observed.  A second gapped lobe centered at $\vert B\vert\sim220$~mT then appeared, containing several subgap states away from zero energy, as shown in greater detail in \cite{Supplementary}. The second lobe closes at $250$~mT, above which only normal-state behavior was observed.

The dependence of tunneling spectra on back-gate voltage in the zeroth lobe is shown in Fig.~2C. In weak tunneling regime, for $V_{\rm BG} < -1$~V a hard gap was observed, with $\Delta = 180~\mu$eV (Fig.~2,~C~and~D). As the device is opened, for $V_{\rm BG} \sim -0.8$~V subgap conductance is enhanced due to Andreev processes. The resonance at $V_{\rm BG} \sim -1.2$~V is likely due to a resonance in the barrier. In the first lobe, at $B=110$~mT, the sweep of $V_{\rm BG}$ showed a zero-energy state throughout the tunneling regime (Fig.~2E). The cut displayed in Fig.~2F shows a discrete zero-bias peak well separated from other states. As the tunnel barrier is opened, the zero bias peak splits and eventually evolves into a zero-bias dip at strong coupling, in qualitative agreement with theory supporting MZMs \cite{Vuik2018}. Additional line-cuts as well as the tunneling spectroscopy for the second lobe are provided in \cite{Supplementary}. Several switches in data occurred at the same gate voltages in Fig.~2,~C~and~E, presumably due to gate-dependent charge motion in the barrier.

\begin{figure}[t]
\includegraphics[width=\linewidth]{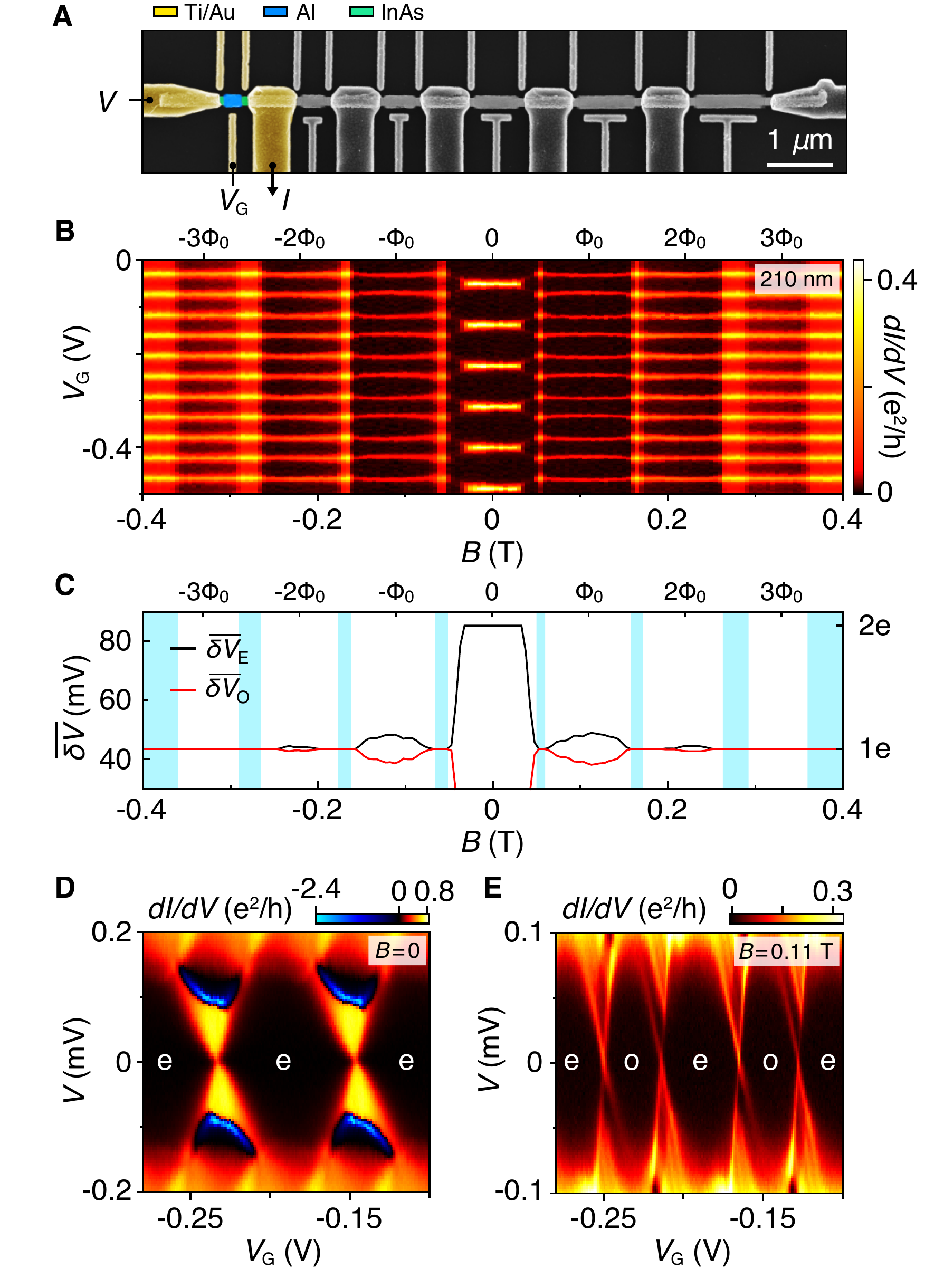}
\caption{\label{fig:3} \textbf{Short-island Coulomb blockade: $\bf 2e$ peaks in the zeroth lobe, even-odd peaks in the first lobe.} (\textbf{A}) Micrograph of device 2 comprising six islands with individual gates and leads, spanning a range of lengths from $210$~nm to $970$~nm. The measurement setup for 210~nm segment is highlighted in colors. (\textbf{B}) Zero-bias conductance for the 210~nm segment showing Coulomb blockade evolution as a function of plunger gate voltage, $V_{\rm G}$, and axial magnetic field, $B$. (\textbf{C}) Average peak spacings for even (black) and odd (red) Coulomb valleys, $\overline{\delta V}$, from the data in (A) as a function of $B$, with destructive regimes shown in blue. Coulomb peaks spaced by $2e$ split in field and become $1e$-periodic around $55$~mT. At higher field, odd Coulomb valleys shrink, reaching a minimum around $120$~mT. In the second destructive regime around 165~mT peaks are $1e$-periodic again. (\textbf{D}) Zero-field conductance as a function of $V$ and $V_{\rm G}$, showing $2e$ Coulomb diamonds with even (e) valleys only. The negative differential conductance is associated with quasiparticle trapping on the island (see text). (\textbf{E}) Similar to (D) but measured in the first lobe at $B = 110$~mT, reveals discrete, near-zero-energy state, even (e) and odd (o) valleys of different sizes, and alternating excited state structure.
}
\end{figure}

Hybridization of MZMs can be measured in Coulomb islands of finite length from the spacing of Coulomb blockade conductance peaks \cite{Albrecht2016,vanHeck2016,OFarell2018,Shen2018}. In particular, the exponential length dependence of hybridization energy supports the Majorana interpretation and further indicates that the MZMs are located close to the ends of the wire, and not in the middle \cite{Liu2017,Moore2018}. We investigated full-shell islands over a range of device lengths from $210$~nm to $970$~nm, fabricated on a single nanowire, as shown in Fig.~3. 

Zero-bias conductance as a function of plunger gate voltage, $V_{\rm G}$, and $B$ for device 2 yielded series of Coulomb blockade peaks for each segment, examples of which are shown in Fig.~3B. The corresponding average peak spacings, $\overline{\delta V}$, for even and odd Coulomb valleys as a function of $B$ are shown in Fig.~3C. Around zero field, Coulomb blockade peaks with $2e$ periodicity were found. These peaks split at $\sim$40~mT toward the high-field end of the zeroth superconducting lobe, as the superconducting gap decreased bellow the charging energy of the island. The peaks then became $1e$-periodic (within experimental sensitivity) around $55$~mT and throughout the first destructive regime. When superconductivity reappeared in the first lobe, the Coulomb peaks did not become spaced by $2e$ again, but instead showed nearly $1e$ spacing with even-odd modulation. Qualitatively similar even-odd spacing was observed in the second lobe. Unlike device 1 described in Fig.~2, the shortest island in device 2 also showed a third superconducting lobe, which can be identified from the peak height contrast in Fig.~3B. Coulomb blockade peaks were $1e$-periodic within experimental sensitivity throughout the third lobe.

Tunneling spectra at finite source-drain bias showed $2e$ Coulomb diamonds around zero field (Fig.~3D) and nearly $1e$ diamonds at $B = 110$~mT, near the middle of the first lobe (Fig.~3E). The zero-field diamonds are indistinguishable from each other, showing a region of negative differential conductance associated with the onset of quasiparticle transport \cite{Hekking1993,Hergenrother1994,Higginbotham2015}. In the first lobe (Fig.~3E), Coulomb diamonds alternate in size and symmetry, with degeneracy points showing sharp, gapped structure, indicating that the near-zero-energy state is discrete. Additional resonances at finite bias reflect excited discrete subgap states away from zero energy.

\begin{figure}[t]
\includegraphics[width=\linewidth]{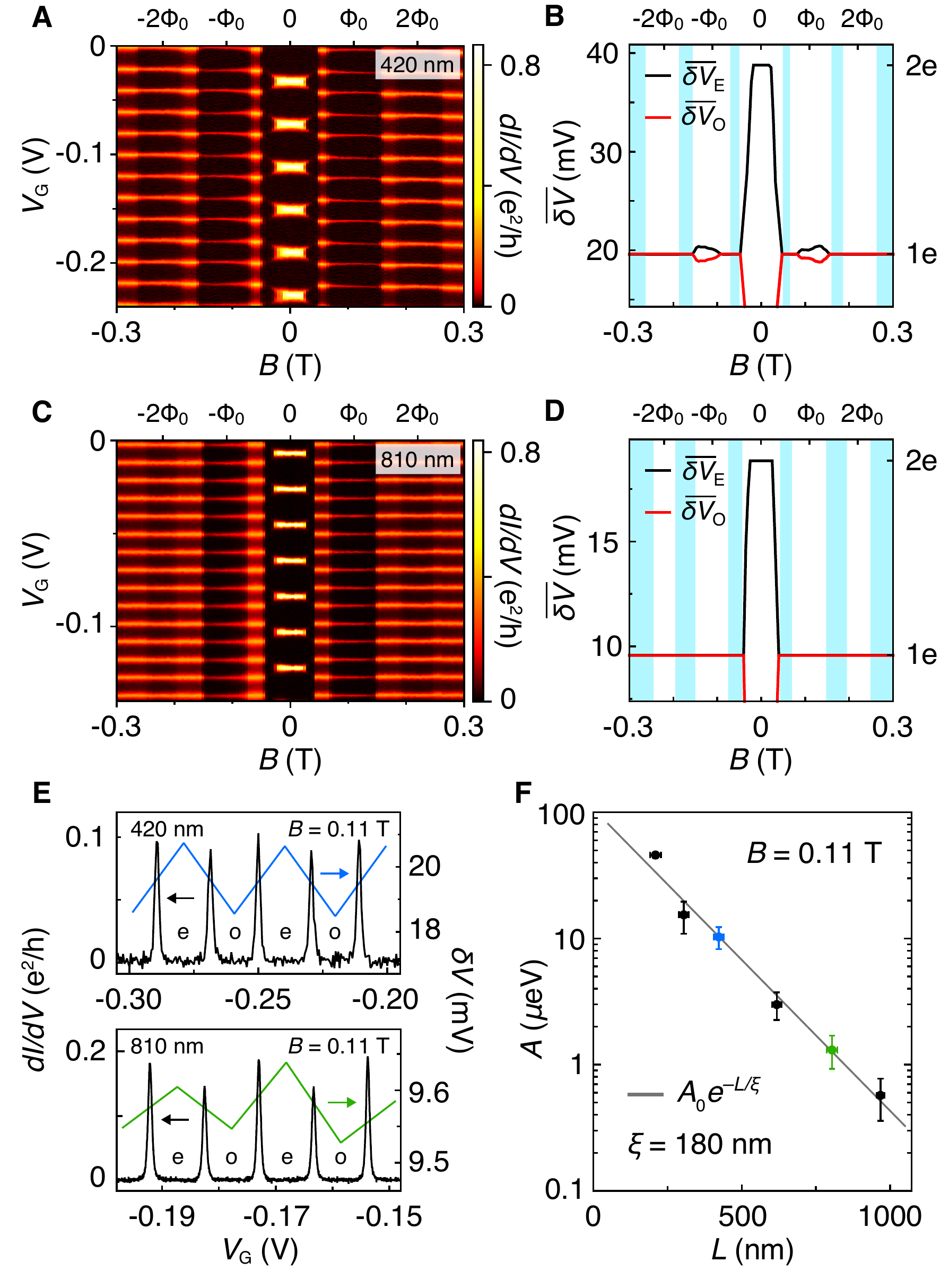}
\caption{\label{fig:4} \textbf{Length dependence of even-odd peak spacing.} (\textbf{A}) Zero-bias conductance showing Coulomb blockade evolution with $V_{\rm G}$ and $B$ for $420$~nm island. (\textbf{B}) Average peak spacing for data in (A). Even-odd pattern is evident in the first lobe, around $B = 110$~mT. (\textbf{C} and \textbf{D}) Similar to (A) and (B) for $810$~nm island. Even-odd spacing in the first lobe is not visible on this scale. (\textbf{E}) Fine-scale Coulomb peak conductance (black, left axis) and spacing (colored, right axis) as a function of plunger gate voltage, $V_{\rm G}$ at $B=110$~mT for $420$~nm and $810$~nm islands. (\textbf{F}) Average even-odd peak spacing difference converted to energy, $A$, using separately measured level arms for each segment, at $B=110$~mT as a function of island length, $L$, along with the best fit to the exponential form $A = A_{\rm 0} e^{-L/\xi}$, giving the best fit parameters $A_{\rm 0}=110~\mu$eV and $\xi=180$~nm. Vertical error bars indicate uncertainties from standard deviation of $\overline{\delta V}$ and lever arms. Experimental noise floor, $\sigma_{A}<0.1\,\mu{\rm eV} \ll k_{\rm B}T$, measured using $1e$ spacing in destructive regime. Horizontal error bars indicate uncertainties in lengths estimated from the micrograph. 
}
\end{figure}

Coulomb peaks for two longer islands are shown in Fig.~4,~A~to~E, with full data sets for other lengths reported in \cite{Supplementary}. All islands showed $2e$-periodic Coulomb peaks in the zeroth lobe and nearly $1e$ spacing in the first lobe. Examining the 420~nm and 810~nm data in Fig.~4,~A,~C~and~E already reveals that the mean difference between even and odd peak spacings in the first lobe decreased with increasing island length. To address this question quantitatively, we determine the lever arm, $\eta$, for each island independently in order to convert plunger gate voltages to chemical potentials on the islands, using the slopes of the Coulomb diamonds \cite{Thijssen2008, Albrecht2016}.  This allows the peak spacing differences (Fig.~4,~B~and~D) to be converted to island-energy differences, $A(L)$, between even and odd occupations, as a function of device length, $L$. Within a Majorana picture, the energy scale $A(L)$ reflects the length dependent hybridization energy of MZMs. Values for $A(L)$ at $B = 110$~mT, in the middle of the first lobe, spanning over two orders of magnitude are shown in Fig.~4F. A fit to an exponential $A = A_{\rm 0} e^{-L/\xi}$ yields fit parameters $A_{\rm 0}=110~\mu$eV and $\xi=180$~nm. The data are well described by an exponential length dependence, implying that the low-energy modes are located at the ends of the wire, not bound to impurities or local potential fluctuations as expected for overlapping Majorana modes. Along with length dependent even-odd peak spacing difference, we observe even-odd modulation in peak heights (Fig.~4E), as described theoretically in Ref.~\cite{Hansen2018}, with a complex alternating structure within the first lobe. Peak height modulation accompanying peak spacing modulation was observed previously \cite{Albrecht2016,OFarell2018,Shen2018}.

We note that $\xi$ from the fit in Fig.~4F matches the coherence length of the Al shell, $\xi_{\rm s}=180$~nm. While it remains unclear if these quantities are necessarily equal, we note that this characteristic scale also emerges by examining spectral features in the first lobe in Fig.~2B. We tentatively identify subgap features as excitations of the trapped vortex, analogous to Caroli-de~Gennes-Matricon (CdGM) states \cite{Caroli1964}, in this case located close to the semiconductor-superconductor interface due to band bending. The first excited state at $B = 110$~mT is visible at $\delta_{110} \sim 50\,\mu$eV in Fig.~2,~B,~E~and~F. From the CdGM relation $\delta = \Delta/(k\xi)$, with an effective wavevector $k$. Taking $\Delta_{110}\sim130\,\mu$eV as the gap at $B=110$~mT and the excitation wavelength to be the boundary circumference, $2\pi/k\sim\pi{\rm D}\sim 400$~nm, yields $\xi\sim(\Delta_{110}/\delta_{110})~{\rm D}/2 \sim 170$~nm.  Within this picture, the interaction between MZMs is mediated by cylindrical electronic modes near the superconductor-semiconductor interface. 

As a check of this interpretation, we reexamine the length dependence of even-odd peak spacing closer to the edge of the first node, at $B=140$~mT, where a reduced gap, $\Delta_{140} = 40\,\mu$eV, with no excited subgap states was found \cite{Supplementary}. The expected scaling of coherence lengths, $\xi_{140}/\xi_{110} = \delta_{110}/\Delta_{140} \sim 5/4$, is in good agreement with the coherence length extracted from the exponential length dependence at 140~mT, $\xi_{140} = 230$~nm, as shown in \cite{Supplementary}. We speculate that the absence of subgap states near the high-field edge of the lobe is related to a breakdown of the CdGM picture as the orbits associated with subgap states for the reduced gap become localized to the semiconductor-superconductor interface. 

In summary we have demonstrated that threading magnetic flux through an InAs nanowire with a fully surrounding epitaxial Al shell can induce a topological phase with Majorana zero modes at the nanowire ends. The modest magnetic field requirements, protection of the semiconductor core from surface defects, and locked phase winding in discrete lobes together suggest a new and relatively easy route to creating and controlling Majorana zero modes in hybrid materials. 

We thank L.~Casparis, K.~Flensberg, A.~Higginbotham, T.~Karzig, R.~M.~Lutchyn, C.~Nayak, B.~van~Heck and G.~W.~Winkler for valuable discussions, as well as C.~S\o rensen, R.~Tanta and S.~Upadhyay for contributions to material growth and device fabrication. Research was supported by Microsoft, the Danish National Research Foundation, and the European Commission. M.T.D.~acknowledges support from State Key Laboratory of High Performance Computing, China.

\onecolumngrid
\clearpage
\onecolumngrid
\setcounter{figure}{0}
\setcounter{equation}{0}
\section{\large{S\MakeLowercase{upplemental} M\MakeLowercase{aterial}
}}
\renewcommand{\figurename}{FIG.~S}
\renewcommand{\tablename}{Table.~S}
\renewcommand{\thetable}{\arabic{table}}
\twocolumngrid

\section{Materials and Methods}

\textbf{Nanowire growth} The hybrid nanowires used in this work were grown by molecular beam epitaxy on InAs(111)B substrate at $420~^\circ$C. The growth was catalyzed by Au via the vapor-liquid-solid method. The nanowire growth was initiated with an axial growth of InAs along the $[0001]$ direction with wurtzite crystal structure, using an In flux corresponding to a planar InAs growth rate of $0.5~\mu$m/hr and a calibrated As$_{4}$/In flux ratio of 14. The InAs nanowires with core $130$~nm were grown to a length of $\sim 10~\mu$m. Subsequently, an Al shell with thickness of $30$~nm was grown at $-30~^\circ$C on all six facets by continuously rotating the growth substrate with respect to the metal source. The resulting full shell had an epitaxial, oxide-free interface between the Al and InAs \cite{supKrogstrup2015}.
 
\textbf{Device fabrication} The devices were fabricated on a degenerately n-doped Si substrate capped with a $200$~nm thermal oxide. Prior to the wire deposition, the fabrication substrate was pre-fabricated with a set of alignment marks as well as bonding pads. Individual hybrid nanowires were transfered from the growth substrate onto the fabrication substrate using a manipulator station with a tungsten needle. Standard electron beam lithography techniques were used to pattern etching windows, contacts and gates. The quality of the Al etching was found to improve when using a thin layer of AR 300-80 (new) adhesion promoter. Double layer of EL6 copolymer resists was used to define the etching windows. The Al was then selectively removed by submerging the fabrication substrate for $60$~s into MF-321 photoresist developer. As the native InAs and Al oxides have different work functions, different cleaning processes had to be applied before contacting the wires. To contact the Al shell in device 1, a stack of A4 and A6 PMMA resist was used. Normal Ti/Al ($5/210$~nm) ohmic contacts to Al shell were deposited after \textit{in-situ} Ar-ion milling (RF ion source, $25$~W, $18$~mTorr, $9$~min). To contact the InAs core in both device 1 and 2, a single layer of A6 PMMA resist was used. A gentler Ar-ion milling (RF ion source, $15$~W, $18$~mTorr, $6.5$~min) was used to clean the InAs core followed by metalization of the normal Ti/Al ($5/180$~nm) ohmic contacts to InAs core. A single layer of A6 PMMA resist was used to form normal Ti/Al ($5/150$~nm) in device 2.

\begin{figure}[!h!b]
\centering
\includegraphics[width=\linewidth]{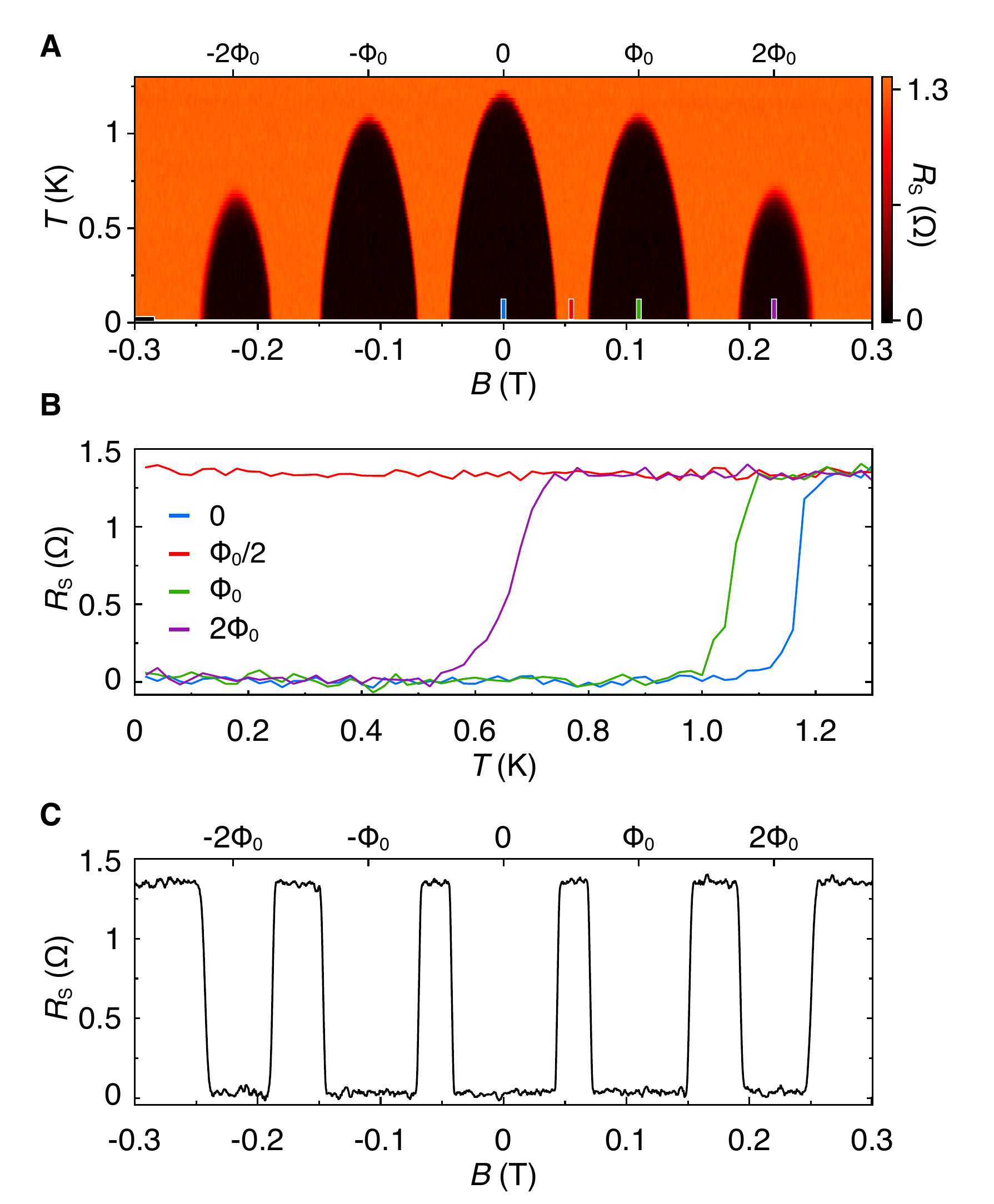}
\caption{\label{fig:S1} \textbf{Shell resistance versus temperature and magnetic field.} (\textbf{A}) Same data as in the main text Fig.~1D: Differential resistance of the Al shell, $R_{\rm s}$, measured for device 1 as a function of temperature, $T$, and axial magnetic field, $B$. (\textbf{B}) Line-cuts from (A) taken at $0$, $1/2$, $1$ and $2$ flux quanta, $\Phi_0$. At half of the flux quantum, $R_{\rm s}$ stays at the normal state resistance down to the lowest measured temperature $T=20$~mK. (\textbf{C}) Line-cut from (A) taken at $T=20$~mK. Two destructive regimes surrounded by fully superconducting phase can be  seen around $\vert B\vert = 55$ and $165$~mT.}
\end{figure}

\textbf{Measurements} Each of the dc lines used to measure and gate the devices was equipped with RF and RC filters (QDevil \cite{supQDevil}), adding a line resistance $R_{\rm Line}=6.7$~k$\Omega$. The 4-probe differential resistance measurements were carried out using an ac excitation of $I_{\rm ac}=200$~nA. The 2-probe tunneling conductance measurements were conducted using ac excitation of $V_{\rm ac}= 5~\mu$V.

\section{Supplementary Text}

\textbf{Destructive regime} As a result of fluxoid quantization, the critical temperature, $T_{\rm C}$, of a cylindrical shell is periodically modulated by an axial magnetic field \cite{supTinkham1966,supLittle1962}. For cylinders with radius smaller than the superconducting coherence length, $T_{\rm C}$ is expected to vanish whenever the applied flux (axial magnetic field component times shell cross-sectional area) is close to an odd half-integer multiple of the superconducting flux quantum, $n\Phi_0/2$ ($\Phi_0 = h/2e = 2.07$~mT$\,\mu$m$^{2}$, $n=1,3,5,\ldots$) \cite{supdeGennes1981,supSchwiete2009,supDao2009}. Throughout the extended range where $T_{\rm C}$ vanishes, superconductivity is destroyed \cite{supLiu2001,supSternfeld2011}.

The measured nanowires have hexagonal InAs core with diameter of $130$~nm and Al shell with thickness of $30$~nm, giving a mean diameter of $\sim 160$~nm. The dirty-limit coherence length is given by $\xi_{\rm S}=\sqrt[]{\pi\hbar v_{\rm F} l/24 k_{\rm B} T_{\rm C}}$ \cite{supTinkham1966,supGordon1984}. The measured normal state resistance is $R_{\rm N}=1.3~\Omega$. The distance between the voltage probes is $\sim 940$~nm. This yields shell resistivity of $\rho_{\rm S} = 21$~nm~$\Omega$. The Fermi velocity in Al is $v_{\rm F}=2\times 10^6$~m/s \cite{supKittel2005}, giving a Drude mean free path of $l=19$~nm. The measured critical temperature is $T_{\rm C}=1.2$~K. This gives $\xi_{\rm S}=180$~nm, greater than the mean nanowire radius ($\sim 80$~nm), hence the measured nanowires are expected to exhibit a destructive regime. This is consistent with the measurements, see Fig.~S1A.

At integer flux quanta, normal-to-superconducting transitions appear as the temperature is lowered, with the critical temperature decreasing as the flux number increases. Around $\pm\Phi_0/2$ and $\pm3\Phi_0/2$\ the resistance of the shell, $R_{\rm s}$, stays at the normal value down to the lowest measured temperature, $\sim20$~mK, as shown in Fig.~S1B. At the base temperature, the two destructive regimes can be identified by abrupt changes of $R_{\rm s}$ from $0$ to $R_{\rm N}$ and then back to $0$ when the flux passes $\pm\Phi_0/2$ and $\pm 3\Phi_0/2$, see Fig.~S1C.

\textbf{Penetration depth} An applied magnetic field penetrates thin-film superconductors with thickness much less than penetration depth, $\lambda$, uniformly. In dirty limit, the effective penetration depth $\lambda_{\rm eff} = \lambda_{\rm L}~\sqrt[]{\xi_0/(1.33~l)}$, where $\xi_0$ and $\xi_{\rm S}$ at zero temperature are related by $\xi_{\rm S} = 0.855~\sqrt[]{\xi_0 l}$ \cite{supTinkham1966}. Taking $\lambda_{\rm L}=16~nm$ as the London penetration depth for Al \cite{supKittel2005}, yields $\lambda_{\rm eff}=150$~nm greater than Al thickness (30~nm). As a result, the flux in the wire is not quantized. Note, however, that the fluxoid is still quantized \cite{supLittle1962,supTinkham1966}.

\begin{figure}[!t]
\centering
\includegraphics[width=\linewidth]{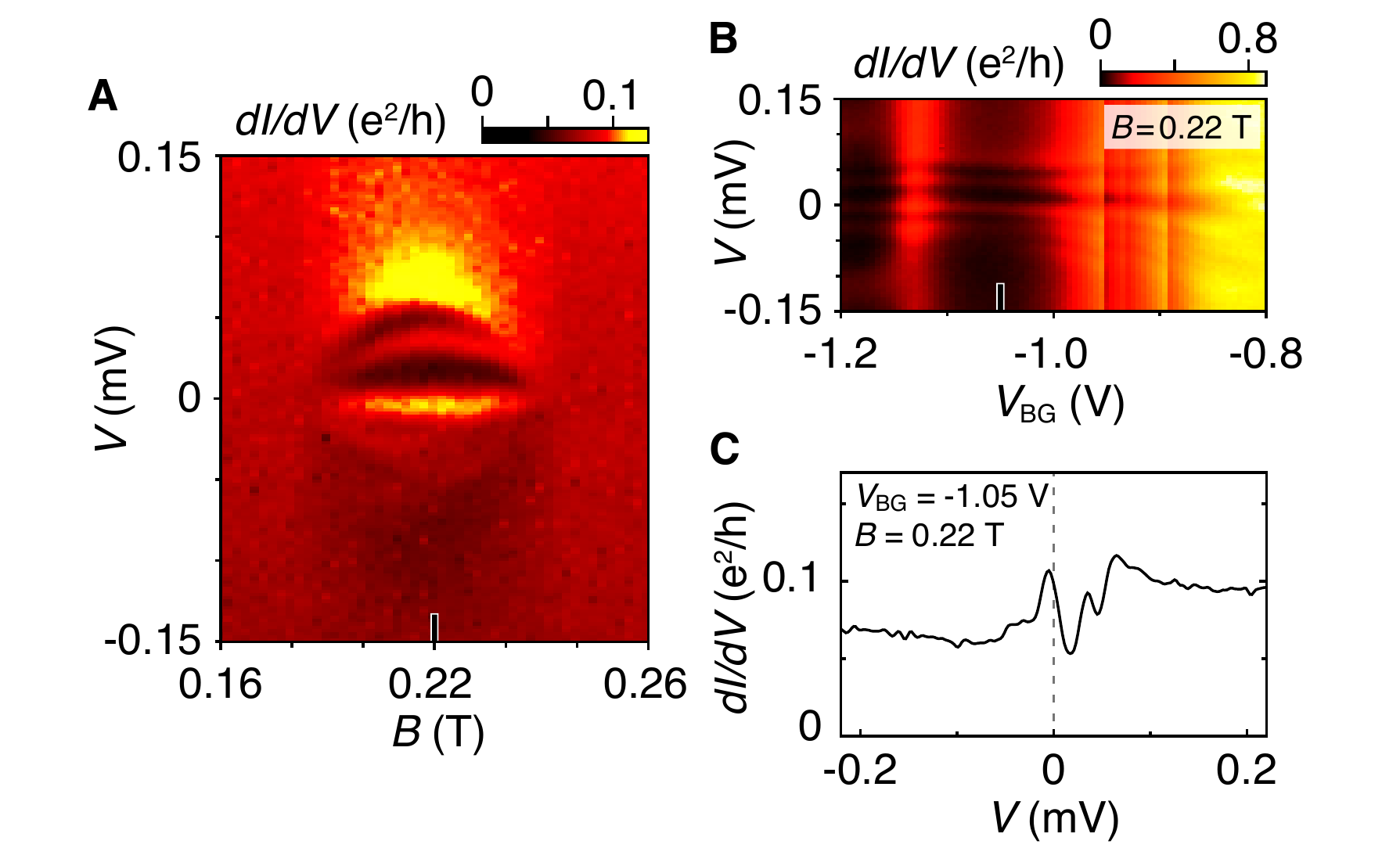}
\caption{\label{fig:S2} \textbf{Tunneling spectroscopy in the second lobe.} (\textbf{A}) Zoom-in around the second superconducting lobe of the data shown in the main text Fig.~2B. (\textbf{B}) Differential conductance as a function of source-drain voltage, $V$, and back-gate voltage, $V_{\rm BG}$. (\textbf{E}) Line-cut of the conductance taken at $B=0.22$~T and $V_{\rm BG}=-1.05$~V. The spectrum shows subgap states away from zero energy.
}
\end{figure}

\textbf{Tunneling spectrum}  The zeroth lobe, where the winding number is 0, shows a hard gap and no subgap states are visible. In the first lobe, with the phase winding of $2\pi$, the spectrum displays a discrete, zero-energy state. In the second lobe, with even number of phase windings, the spectrum features an asymmetric superconducting density of states with the lowest energy subgap state centered around $\sim -5~\mu$eV, see Fig.~S2. Note that Majorana zero modes are expected to appear only when the superconducting phase winding number is odd \cite{supLutchyn2018_2}. The energy of the state does not depend on the coupling strength to the probe, set by back-gate voltage, $V_{\rm BG}$. No sign of the third lobe is detected for device 1.

\textbf{Coulomb spectroscopy} The Coulomb peak spacing is dictated by the lowest energy state at energy $E_0$, may it be a subgap state or the superconducting gap itself. The periodicity of the Coulomb peaks is determined by the ratio between $E_0$ and the charging energy, $E_{\rm C}$. The Coulomb blockade is $2e$ periodic for $E_0>E_{\rm C}$; It becomes even-odd once $E_0$ is less than $E_{\rm C}$; And it is $1e$ periodic in case $E_0 = 0$. Non-interacting Majorana modes have zero energy, hence a Coulomb island hosting Majoranas can be charged in portions of single electrons. If the wavefunctions of the opposing Majorana modes have a finite overlap, for example because of the finite island length, the energy of the corresponding modes will deviate from zero \cite{supAlbrecht2016,supvanHeck2016}.

\onecolumngrid
\begin{center}
\begin{figure}[!b!h]
\centering
\includegraphics[width=\linewidth]{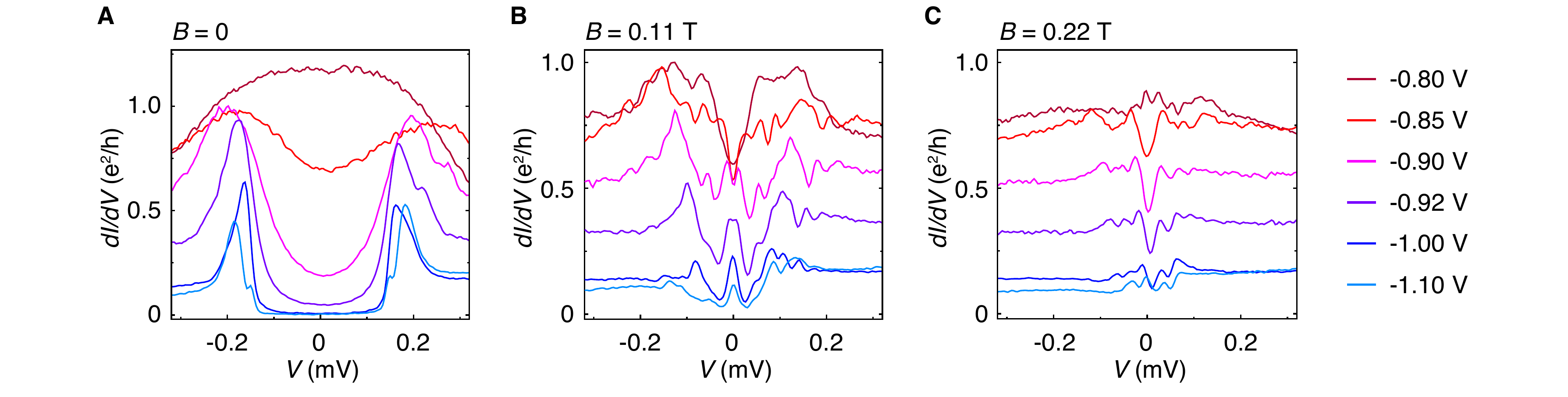}
\caption{\label{fig:S3} \textbf{Spectrum evolution with barrier strength.} Line-cuts of the conductance at different back-gate voltages, $V_{\rm BG}$, measured for device 1 at (\textbf{A}) $B=0$, around zero flux, (\textbf{B}) $B=0.11$~T, around one flux quantum, and (\textbf{C}) $B=0.22$~T, around two flux quanta.
}
\end{figure}
\end{center}
\twocolumngrid

\begin{figure}[!h!t]
\centering
\includegraphics[width=1\linewidth]{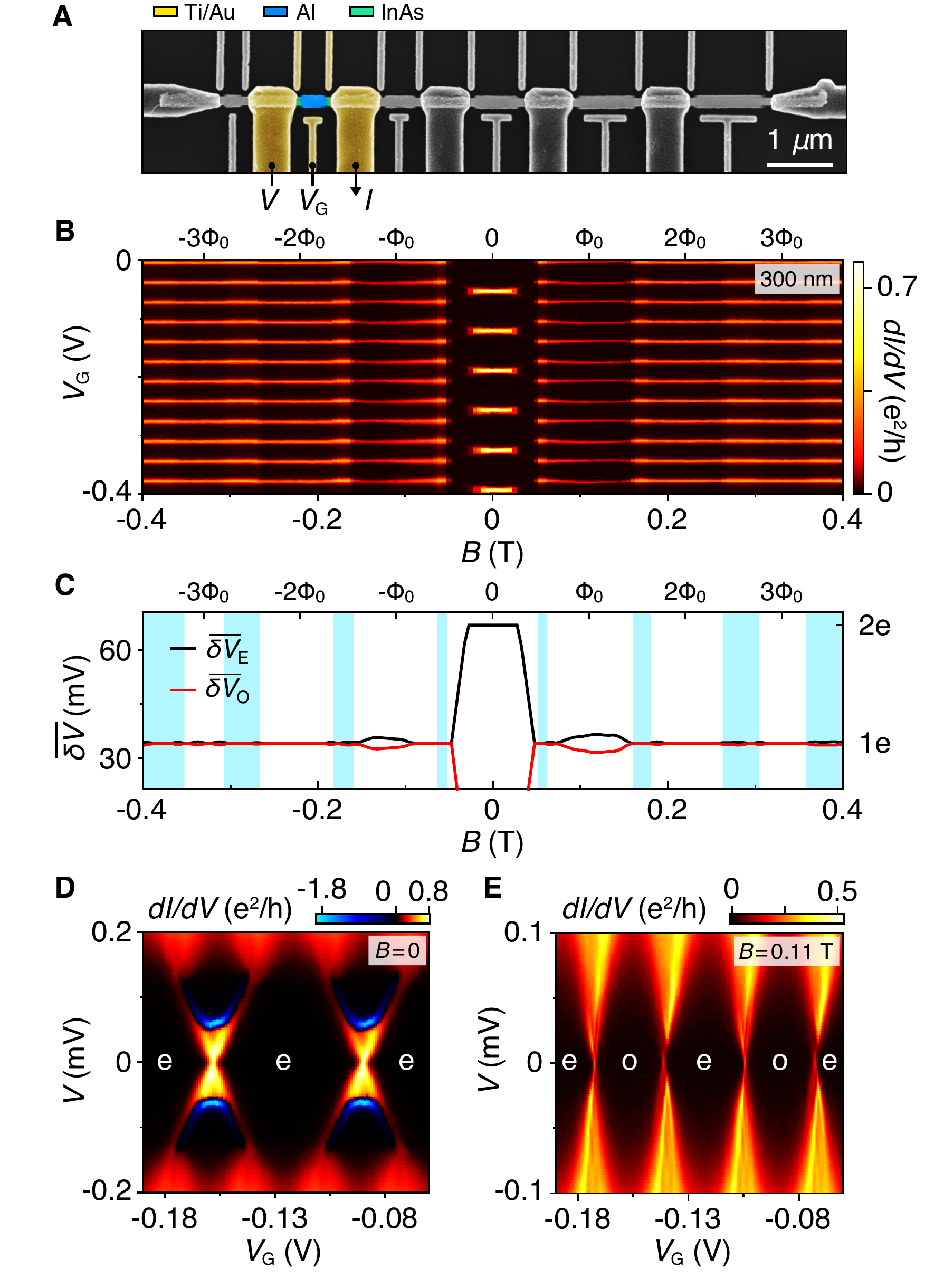}
\caption{\label{fig:S4} \textbf{300 nm Coulomb island.}
(\textbf{A}) Micrograph of device 2 with the measurement setup for $300$~nm island highlighted in colors. (\textbf{B}) Zero-bias conductance showing Coulomb blockade evolution as a function of plunger gate voltage, $V_{\rm G}$, and magnetic field, $B$. (\textbf{C}) Average peak spacing for even (black) and odd (red) Coulomb valleys, $\overline{\delta V}$, from the measurements shown in (A) as a function of $B$. The blue background indicates the magnetic field ranges where superconductivity is absent. (\textbf{D}) Zero-field conductance as a function of $V$ and $V_{\rm G}$.(\textbf{E}) Similar to (D) but measured at $B = 110$~mT.}
\end{figure}

\begin{figure}[!h!t]
\centering
\includegraphics[width=\linewidth]{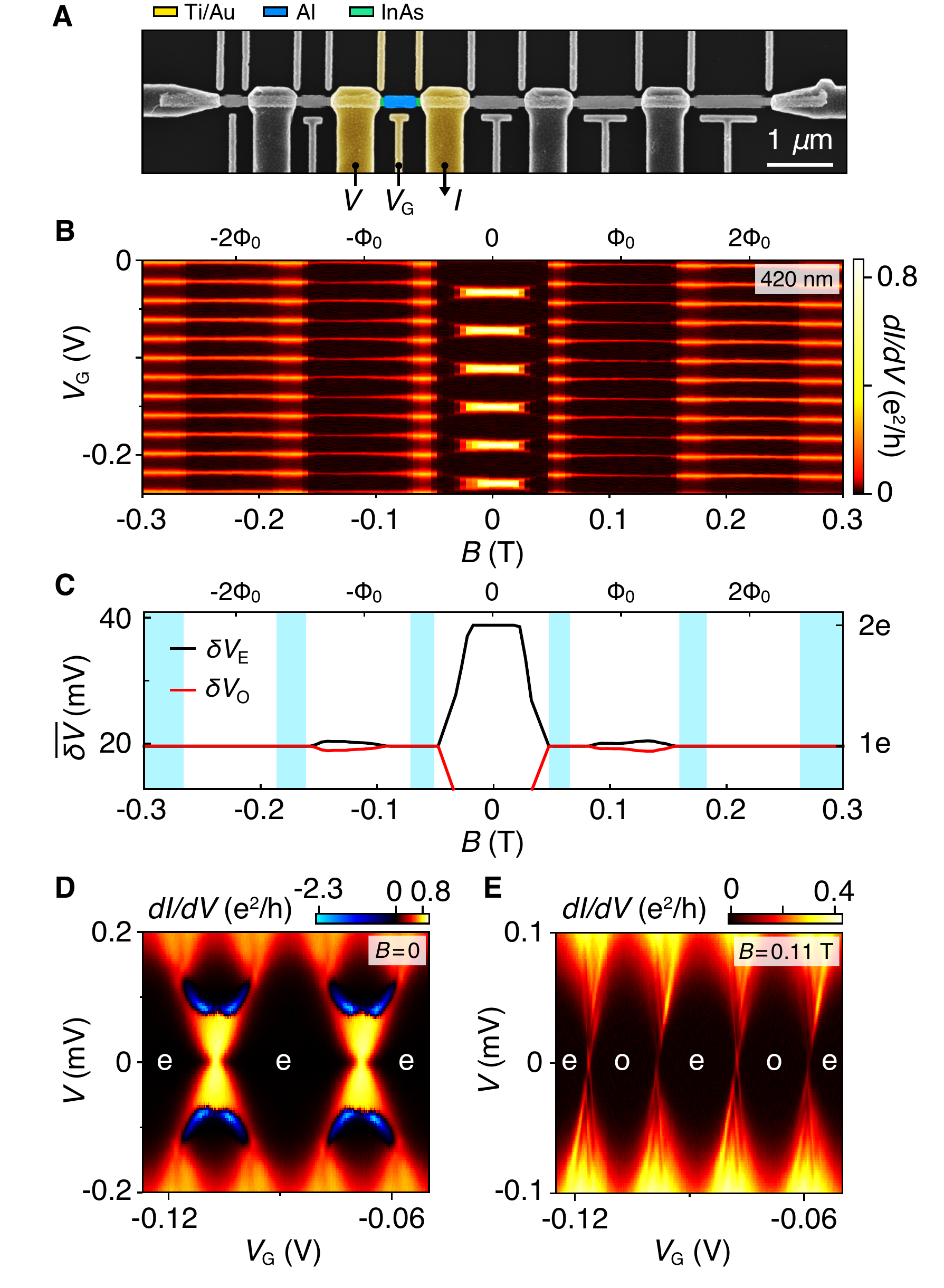}
\caption{\label{fig:S5} \textbf{420 nm Coulomb island.}
(\textbf{A}) Micrograph of device 2 with the measurement setup for $420$~nm island highlighted in colors. (\textbf{B}) Zero-bias conductance showing Coulomb blockade evolution as a function of plunger gate voltage, $V_{\rm G}$, and magnetic field, $B$. (\textbf{C}) Average peak spacing for even (black) and odd (red) Coulomb valleys, $\overline{\delta V}$, from the measurements shown in (A) as a function of $B$. The blue background indicates the magnetic field ranges where superconductivity is absent. (\textbf{D}) Zero-field conductance as a function of $V$ and $V_{\rm G}$. (\textbf{E}) Similar to (D) but measured at $B = 110$~mT.}
\end{figure}

In the even-odd Coulomb blockade regime, the Coulomb-peak spacing, $\delta V$, is proportional to $E_{\rm C} + 2E_0$ for even diamonds and $E_{\rm C} - 2E_0$ for odd diamonds, which implies that $\delta V_{\rm E} - \delta V_{\rm O} \propto E_0$  \cite{supAlbrecht2016,supHigginbotham2015}. This makes the Coulomb spectroscopy a powerful tool to study the interaction of Majorana modes in quantum dots with finite size.

Device 2 consists of six hybrid quantum dots with lengths $L$ ranging from $210$~nm up to $970$~nm. Figure 3 in the main text presents measurements for the shortest island. Data for the other five islands are presented in Figs.~S4--S8. In each of the figure, panel A displays the scanning electron micrograph with the measurement setup for corresponding island highlighted in false colors; Panel B shows zero-bias conductance as a function of the axial magnetic field, $B$, and gate voltage, $V_{\rm G}$; Panel C depicts average even and odd peak spacing evolution in magnetic field, extracted from the data shown in panel B; Panels D and E show Coulomb diamonds in the middle of the zeroth and first lobes, the later featuring zero-bias peaks at the degeneracy points for each island.

The same measurement routine was carried out at several different gate configurations for each island to gather more statistics. The average lever arm, $\overline{\eta}$, average even and odd peak spacing difference $\Delta\overline{\delta V}_{110}$ as well as the corresponding amplitude $A=\overline{\eta}~\times~\Delta\overline{\delta V}_{110}$---all measured at $110$~mT---are given in Table~S1.

\newpage
\begin{figure}[!h!t]
\centering
\includegraphics[width=\linewidth]{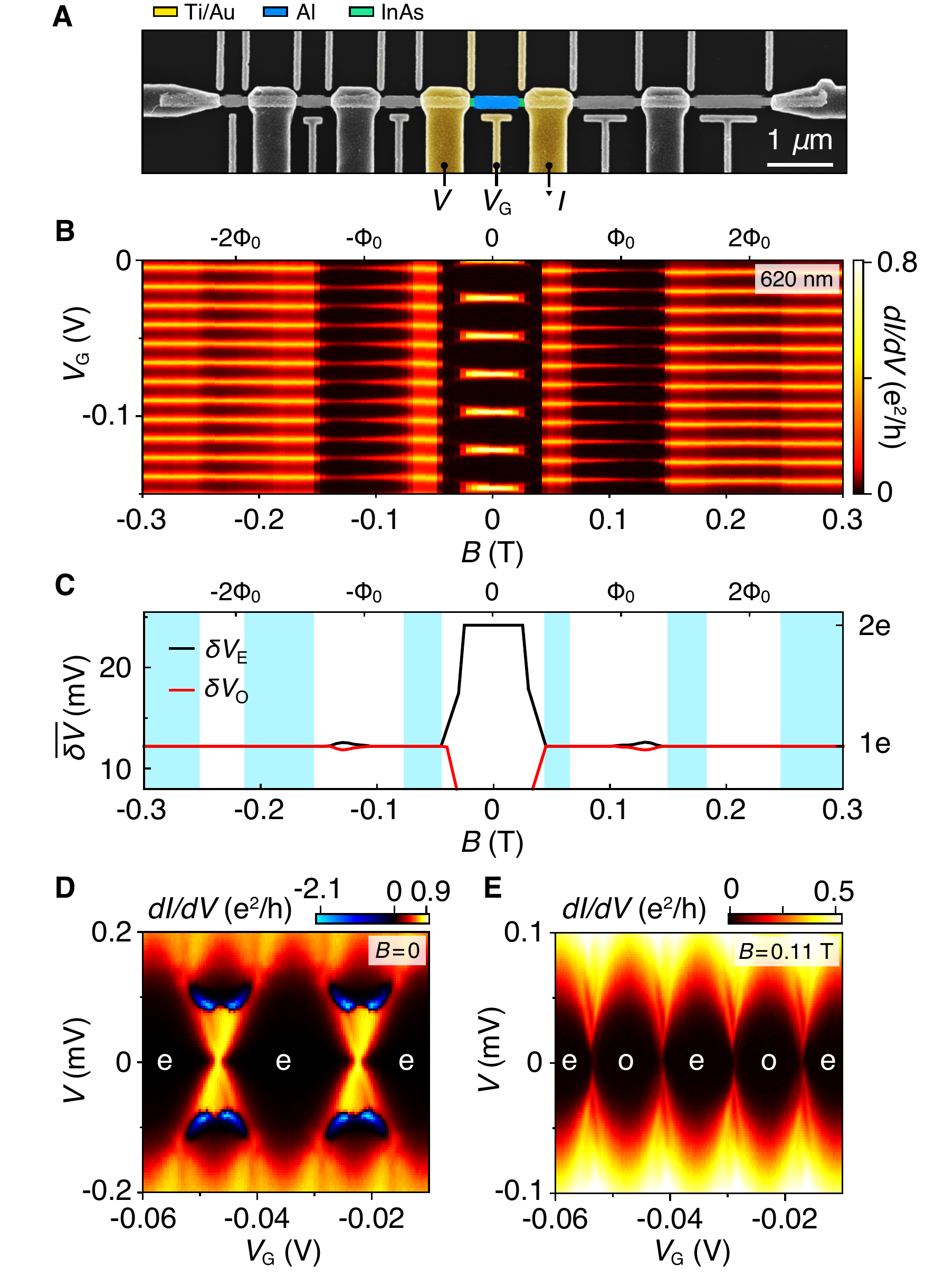}
\caption{\label{fig:S6} \textbf{620 nm Coulomb island.}
(\textbf{A}) Micrograph of device 2 with the measurement setup for $620$~nm island highlighted in colors. (\textbf{B}) Zero-bias conductance showing Coulomb blockade evolution as a function of plunger gate voltage, $V_{\rm G}$, and magnetic field, $B$. (\textbf{C}) Average peak spacing for even (black) and odd (red) Coulomb valleys, $\overline{\delta V}$, from the measurements shown in (A) as a function of $B$. The blue background indicates the magnetic field ranges where superconductivity is absent. (\textbf{D}) Zero-field conductance as a function of $V$ and $V_{\rm G}$. (\textbf{E}) Similar to (D) but measured at $B = 110$~mT.}
\end{figure}

\newpage
\begin{figure}[!h!t]
\centering
\includegraphics[width=\linewidth]{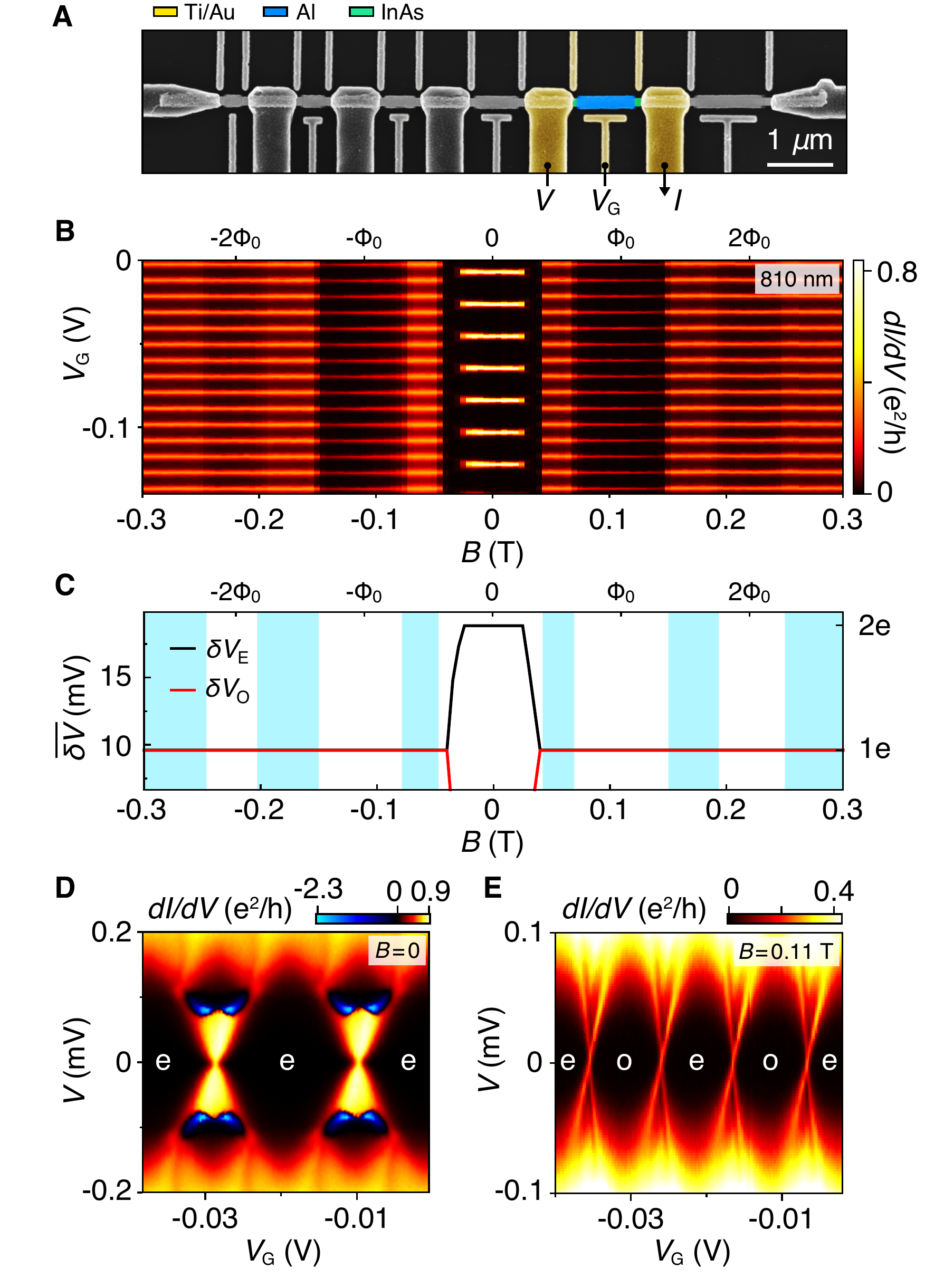}
\caption{\label{fig:S7} \textbf{810 nm Coulomb island.}
(\textbf{A}) Micrograph of device 2 with the measurement setup for $810$~nm island highlighted in colors. (\textbf{B}) Zero-bias conductance showing Coulomb blockade evolution as a function of plunger gate voltage, $V_{\rm G}$, and magnetic field, $B$. (\textbf{C}) Average peak spacing for even (black) and odd (red) Coulomb valleys, $\overline{\delta V}$, from the measurements shown in (A) as a function of $B$. The blue background indicates the magnetic field ranges where superconductivity is absent. (\textbf{D}) Zero-field conductance as a function of $V$ and $V_{\rm G}$. (\textbf{E}) Similar to (D) but measured at $B = 110$~mT.}
\end{figure}

\begin{figure}[!h!t]
\centering
\includegraphics[width=\linewidth]{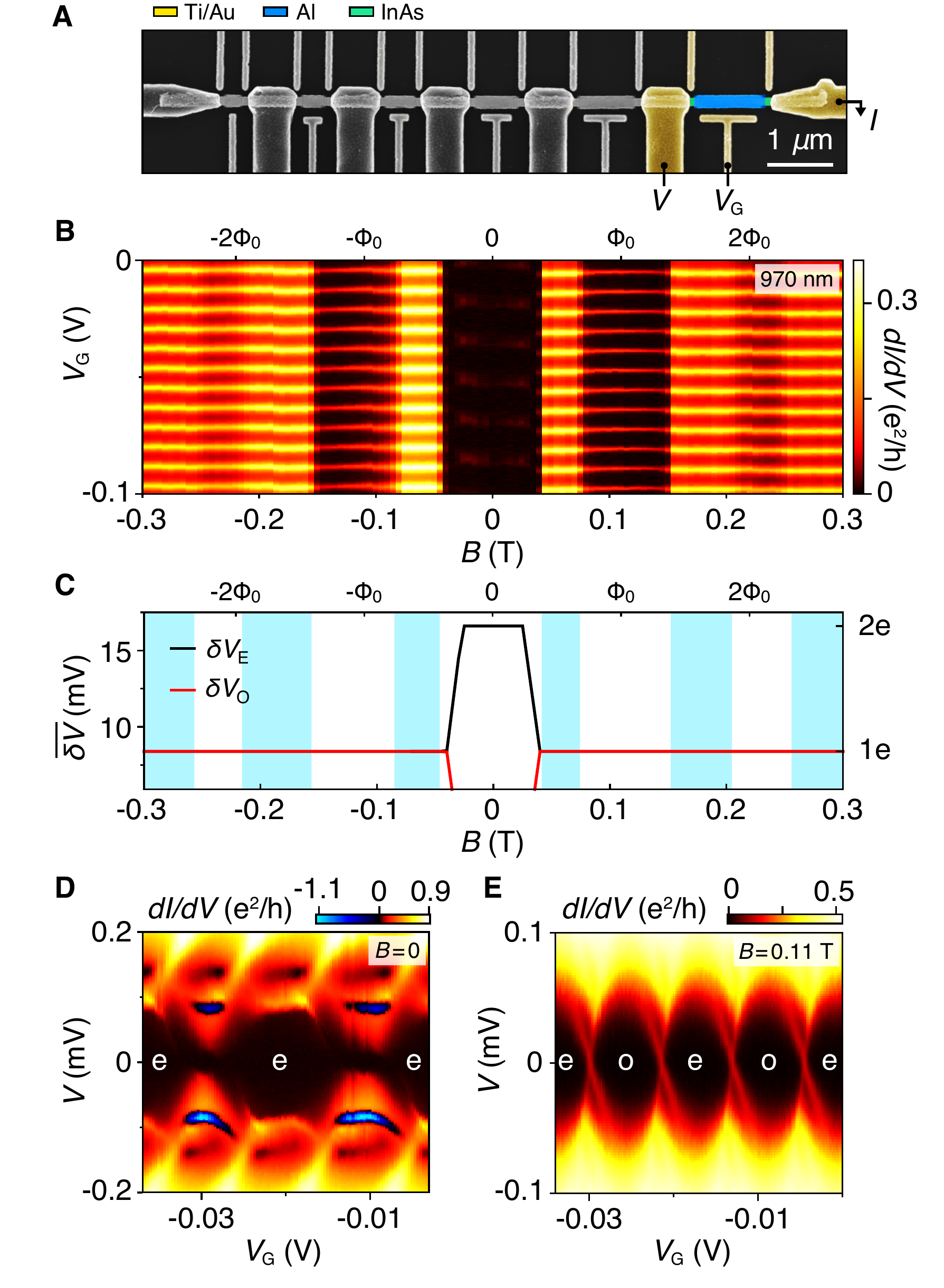}
\caption{\label{fig:S8} \textbf{970 nm Coulomb island.}
(\textbf{A}) Micrograph of device 2 with the measurement setup for $970$~nm island highlighted in colors. (\textbf{B}) Zero-bias conductance showing Coulomb blockade evolution as a function of plunger gate voltage, $V_{\rm G}$, and magnetic field, $B$. (\textbf{C}) Average peak spacing for even (black) and odd (red) Coulomb valleys, $\overline{\delta V}$, from the measurements shown in (A) as a function of $B$. The blue background indicates the magnetic field ranges where superconductivity is absent. (\textbf{D}) Zero-field conductance as a function of $V$ and $V_{\rm G}$. (\textbf{E}) Similar to (D) but measured at $B = 110$~mT.}
\end{figure}

\newpage
\begin{table}[!h]
\centering
\begin{tabular*}{0.9\linewidth}{@{\extracolsep{\fill}}cccc}
\hline\hline
$L$ (nm)&$\overline{\eta}$ (meV/V)&$\Delta\overline{\delta V}_{110}$ (mV)&$A$ ($\mu$eV)\\
\hline
210 & 4.9 & 9.3 & 45\\
300 & 6.1 & 2.5 & 15\\
420 & 11 & 0.91 & 10\\
620 & 17 & 0.17 & 3\\
810 & 17 & 0.08 & 1.3\\
970 & 15 & 0.04 & 0.6\\
\hline\hline
\end{tabular*}
\caption{\label{tab:tableS1} \textbf{Parameters for device 2.} $L$ is the length of the island. $\overline{\eta}$ is the average lever arm extracted from slopes of the Coulomb diamonds measured at $110$~mT. $\Delta(\overline{\delta V}_{110})$ is the even and odd peak spacing differences measured at $110$~mT. $A=\overline{\eta}~\times~\Delta\overline{\delta V}_{110}$ is the corresponding amplitude in energy.}
\end{table}

\newpage
\onecolumngrid
\begin{center}
\begin{figure}[!t!h]
\centering
\includegraphics[width=\linewidth]{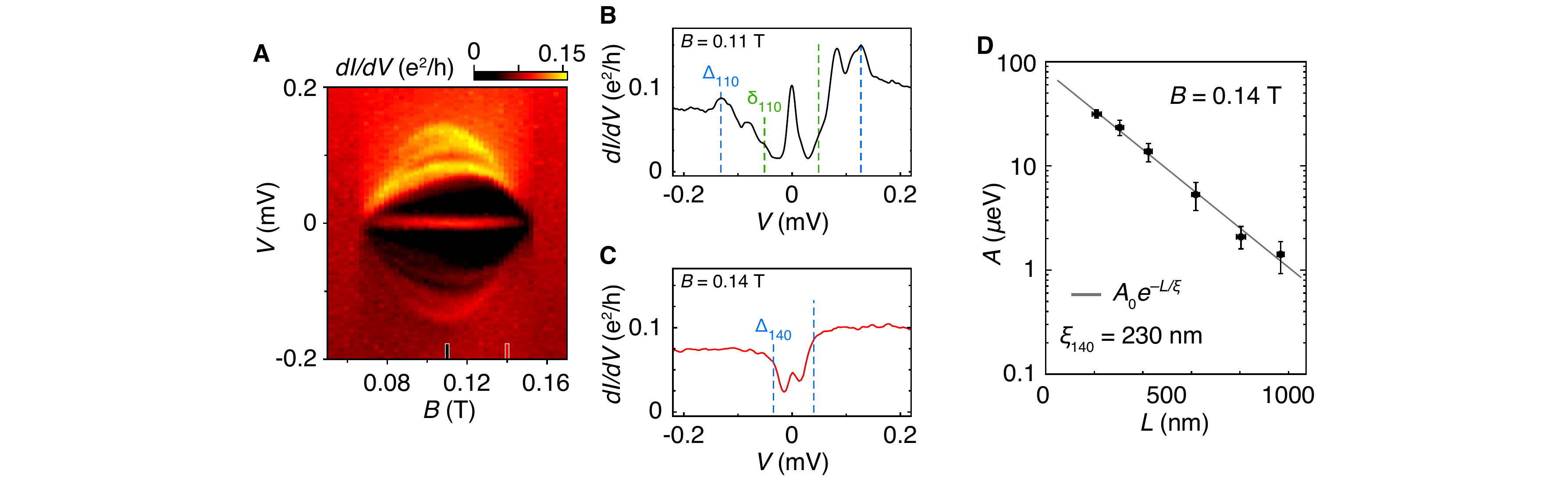}
\caption{\label{fig:S9} \textbf{Superconducting gap and coherence length in the first lobe.}
(\textbf{A}) Zoom-in around the first superconducting lobe of the data shown in the main-text Fig.~2B. (\textbf{B}) Line-cut of the conductance taken from (A) at $B = 110$~mT. The dashed lines indicate the main superconducting gap at $\Delta_{110}=\pm 130~\mu$eV (blue) and the lowest excited state at $\delta_{110}=\pm 50~\mu$eV (green). (\textbf{C}) Line-cut of the conductance taken from (A) at $B = 140$~mT. The blue dashed lines indicate the gap at 140~mT,  $\Delta_{140}=\pm 40~\mu$eV. No subgap states are observed at 140~mT. (\textbf{D}) Average even and odd Coulomb peak spacing difference, $A$, measured at $B = 140$~mT as a function of island length, $L$. The gray line is the best fit to the exponential $A = A_0 e^{-L/\xi}$, yielding $A_0 = 80~\mu$eV and $\xi = 230$~nm. Vertical error bars indicate uncertainties from standard deviation of $\overline{\delta V}$ and lever-arm measured at different gate configurations. Horizontal error bars indicate uncertainties in lengths estimated from the electron micrograph.}
\end{figure}
\end{center}

\newpage

\twocolumngrid

\end{document}